\documentstyle[12pt]{article}
\newcommand{\be}{\begin{equation}}
\newcommand{\ee}{\end{equation}\noindent}
\newcommand{\bear}{\begin{eqnarray}}
\newcommand{\ear}{\end{eqnarray}\noindent}
\newcommand{\no}{\noindent}
\date{}
\renewcommand{\theequation}{\arabic{section}.\arabic{equation}}

\newcommand{\slD}{\raise.15ex\hbox{$/$}\kern-.57em\hbox{$D$}}
\newcommand{\slpartial}{\raise.15ex\hbox{$/$}\kern-.57em\hbox{$\partial$}}
\newcommand{\slG}{{{\dot G}\!\!\!\! \raise.15ex\hbox {/}}}

\def\non{\nonumber}
\def\beqn*{\begin{eqnarray*}}
\def\eqn*{\end{eqnarray*}}

\def\square{\kern1pt\vbox{\hrule height 1.2pt\hbox{\vrule width 1.2pt
   \hskip 3pt\vbox{\vskip 6pt}\hskip 3pt\vrule width 0.6pt}
   \hrule height 0.6pt}\kern1pt}
\def\slash#1{#1\!\!\!\raise.15ex\hbox {/}}

\def\dps{\displaystyle}

\def\half{{1\over 2}}

\def\e{\mbox{e}}

\def\kinb{{1\over 4}\dot x^2}

\def\4piTD{{(4\pi T)}^{-{D\over 2}}}
\def\4piT4{{(4\pi T)}^{-2}}

\def\Tintm4{{\dps\int_{0}^{\infty}}{dT\over T}\,e^{-m^2T}
    {(4\pi T)}^{-2}}
\def\Tintm{{\dps\int_{0}^{\infty}}{dT\over T}\,e^{-m^2T}}

\def\Dx{\dps\int{\cal D}x}

\def\Dpsi{\dps\int{\cal D}\psi}

\begin{document}
\vskip20pt
\hskip12cm LAPTH-687/98\\
\vskip1pt
\begin{center}
\vfill
\large\bf{On the Calculation of QED Amplitudes}\\
\large\bf{in a Constant Field
\footnote{Talk given at the Workshop on ``Frontier Tests of
QED and Physics of the Vacuum'' at Sandansky, Bulgaria, June 9-15,
1998.}
}
\end{center}
\vfill
\begin{center}
Christian Schubert
\footnote{E-mail: schubert@lapp.in2p3.fr}
\\
Laboratoire d'Annecy-le-Vieux
de Physique de Particules\\
Chemin de Bellevue\\
BP 110\\
F-74941 Annecy-le-Vieux CEDEX\\
FRANCE
\end{center}
\vskip.9cm

{\large\bf Abstract}

\begin{quotation}

\noindent
It is explained how first-quantized worldline path integrals can
be used as an efficient alternative to Feynman diagrams in
the calculation of QED amplitudes and effective actions.
The examples include the one-loop photon splitting amplitude,
the two-loop Euler-Heisenberg Lagrangian,
and the one-loop axial vacuum polarization tensor in a
general constant electromagnetic field.
\end{quotation}
\noindent

\vfill

\eject
\pagestyle{plain}
 \setcounter{page}{1}

\section{Introduction}
\renewcommand{\theequation}{1.\arabic{equation}}
\setcounter{equation}{0}

During the last few years it has been established that
techniques originally developed in string perturbation
theory can be used to improve on the
efficiency of calculations in ordinary 
perturbative quantum field theory.
This fact was first demonstrated in the calculation of
tree-level and one-loop QCD scattering amplitudes
\cite{berkos}
but also for one-loop gravity \cite{bkgrav} and 
higher-loop supergravity amplitudes \cite{bddpr}.
Those advantages are due to the superior
organisation of string theory as compared to field 
theory amplitudes. To some extent they can be
recaptured in a more elementary approach \cite{strassler}
based on the representation of one-loop effective
actions in terms of first-quantized path integrals
\cite{feynman,fradkin},
which one evaluates by a technique analogous to the one
which is used for the calculation of the
string path integral.
This approach was found to be particularly well-suited
to the calculation of amplitudes and effective actions
in QED
\cite{strassler,ss1,ss3}.
Here it has been successfully used for calculations
involving constant external fields
\cite{cadhdu,gussho,shais,adlsch,rss,ks},
and also as an easy alternative route to the
construction of multiloop Bern-Kosower type
formulas \cite{ss3}. 

In the present report we focus on the 
constant external field case, and present the following
calculations:
i) One-loop photon splitting in a constant
magnetic field \cite{adlsch}; ii) The two-loop 
Euler-Heisenberg Lagrangian \cite{rss,ks};
iii) The axial vacuum polarization tensor in a general
constant external field. 

\section[]{The Worldline Path Integral Technique in QED}
\renewcommand{\theequation}{2.\arabic{equation}}
\setcounter{equation}{0}

Our starting point for QED calculations is the
following path integral representing
the one - loop effective action for the Maxwell field
due to an electron loop:

\begin{equation}
\Gamma\lbrack A\rbrack = \!-\!2\!\int_0^{\infty}\!
{dT\over T} e^{-m^2T} \!\int \!{\cal D} x {\cal D}\psi\,
{\rm exp}\biggl[\!-\!\!\int_0^T \!\!\!\!\!d\tau
\Bigl({1\over 4}{\dot x}^2 
\!+\! {1\over2}\psi\dot\psi
\!+\! ieA_{\mu}\dot x^{\mu} 
\!-\! ie \psi^{\mu}F_{\mu\nu}\psi^{\nu} \Bigr)\!\biggr]
\label{spinorpi}
\end{equation}

\noindent
Here $T$ is the usual Schwinger proper--time parameter,
the $x^{\mu}(\tau )$'s are the
periodic functions from the circle
with circumference $T$ into $D\;$-- dimensional
spacetime,
and the $\psi^{\mu}(\tau )$'s their
antiperiodic Grassmannian supersymmetric partners.
The analogous formula for scalar QED is obtained simply
by discarding the global factor of $-2$ (which is
for statistics and degrees of freedom) and the Grassman
path integral (which represents the electron spin).
In the ``string--inspired'' approach,
the path integrals over $y$ and $\psi$ 
are evaluated by 
one-dimensional perturbation theory, using the 
Green functions 
\begin{eqnarray}
\langle y^{\mu}(\tau_1)y^{\nu}(\tau_2)\rangle
   &=& - g^{\mu\nu}G_B(\tau_1,\tau_2)
= - g^{\mu\nu}\biggl[\mid\!\tau_1\!-\!\tau_2\!\mid-
{{(\tau_1\!-\!\tau_2)}^2\over T}\biggr]\nonumber\\
\langle \psi^{\mu}(\tau_1)\psi^{\nu}(\tau_2)\rangle
   &=&{1\over 2}\, g^{\mu\nu} G_F(\tau_1,\tau_2)
= {1\over 2}\, g^{\mu\nu}{\rm sign}
(\tau_1 -\tau_2 )
\label{Green}
\end{eqnarray}
\noindent
We 
abbreviate $G_B(\tau_1,\tau_2)=: G_{B12}$ etc.
The bosonic Wick contraction is 
actually carried out in the relative coordinate
$y(\tau)=x(\tau)-x_0$ of the closed loop, while
the (ordinary) integration over the average
position 
$
x_0={1\over T}\int_0^T\!d\tau\, x(\tau)
$
yields energy--momentum conservation.
The result of this evaluation is the 
one-loop effective Lagrangian
${\cal L}(x_0)$. 

One--loop scattering amplitudes are obtained
by specializing the background to 
a finite sum of plane waves of definite
polarization. In the case of scalar QED this
leads to the following 
extremely compact
``Bern-Kosower master
formula'' for the one-loop (off-shell)
N-photon amplitude,

\begin{eqnarray}
\Gamma[k_1,\varepsilon_1;\ldots;k_N,\varepsilon_N]
&=&
{(-ie)}^N
{(2\pi )}^D\delta (\sum k_i)
{\dps\int_{0}^{\infty}}{dT\over T}
{[4\pi T]}^{-{D\over 2}}
e^{-m^2T}
\prod_{i=1}^N \int_0^T 
d\tau_i
\nonumber\\
&&
\!\!\!\!\!\!\!\!\!
\!\!\!\!\!\!\!\!\!\!\!\!\!\!\!\!\!\!\!\!\!\!\!\!\!
\!\!\!\!\!\!\!\!\!\!\!\!\!\!\!\!\!\!\!\!\!\!\!\!\!
\times
\exp\biggl\lbrace \sum_{i,j=1}^N 
\bigl\lbrack \half G_{Bij} k_i\cdot k_j
-i\dot G_{Bij}\varepsilon_i\cdot k_j
+\half\ddot G_{Bij}\varepsilon_i\cdot\varepsilon_j
\bigr\rbrack\biggr\rbrace
\mid_{\rm multi-linear}
\label{scalarqedmaster}
\end{eqnarray}
\no
Here it is understood that only the terms linear
in all the $\varepsilon_1,\ldots,\varepsilon_N$
have to be taken. 
Besides the Green's function $G_B$ also its first and
second deriatives appear,
$\dot G_B(\tau_1,\tau_2) = {\rm sign}(\tau_1 - \tau_2)
- 2 {{(\tau_1 - \tau_2)}\over T},
\ddot G_B(\tau_1,\tau_2)
= 2 {\delta}(\tau_1 - \tau_2)
- {2\over T}
$.
Dots generally denote a
derivative acting on the first variable,
$\dot G_B(\tau_1,\tau_2) \equiv {\partial\over
{\partial\tau_1}}G_B(\tau_1,\tau_2)$, 
and we abbreviate
$G_{Bij}\equiv G_B(\tau_i,\tau_j)$ etc.
The factor ${[4\pi T]}^{-{D\over 2}}$
represents the free Gaussian path integral
determinant factor.

For the fermion QED case an analogous formula can be
written using a superfield formalism \cite{ss3,rss}.
Alternatively the additional 
terms from the Grassmann path integration
can also be generated
by performing a certain partial integration
algorithm on the above expression, and then
applying a simple ``substitution rule'' on the
result \cite{berkos}. As an additional benefit of this
procedure one obtains a permutation symmetric
gauge invariant decomposition of the $N$ - photon
(scalar and fermion QED) amplitudes \cite{Nphoton}.  

\section[]{QED in a Constant Field (1 - Loop)}
\renewcommand{\theequation}{3.\arabic{equation}}
\setcounter{equation}{0}

The presence of an additional constant external field,
taken in Fock-Schwinger gauge centered at $x_0$
\cite{ss1},
changes the path integral Lagrangian
in eq.(\ref{spinorpi}) only by a term
quadratic in the fields,
$
\Delta{\cal L} = {1\over 2}\,ie\,y^{\mu} F_{\mu\nu}
\dot y^{\nu} - ie\,\psi^{\mu} F_{\mu\nu}\psi^{\nu}
$.
The field can therefore be absorbed by a change of the
free worldline propagators, replacing 
$G_B,G_F$ by \cite{rss}

\bear
{{\cal G}_B}(\tau_1,\tau_2) &=&
{1\over 2{(eF)}^2}
\biggl({eF\over{{\rm sin}(eTF)}}
{\rm e}^{-ieTF\dot G_{B12}}
+ ie\,F\dot G_{B12} - {1\over T}\biggr)
\label{calGB}\\
{\cal G}_{F}(\tau_1,\tau_2) &=&
G_{F12} {{\rm e}^{-ieTF\dot G_{B12}}\over {\rm cos}(eTF)}
\label{calGF}
\ear
Thus eq.(\ref{scalarqedmaster}) generalizes to the case
of the scalar QED $N$ - photon scattering amplitude 
in a constant field as follows,

\begin{eqnarray}
\Gamma[k_1,\varepsilon_1;\ldots;k_N,\varepsilon_N]
\!\!\!
&=&
\!\!\!
{(-ie)}^N
{(2\pi )}^D\delta (\sum k_i)
{\dps\int_{0}^{\infty}}{dT\over T}
{[4\pi T]}^{-{D\over 2}}
e^{-m^2T}
{\rm det}^{-{1\over 2}}
\biggl[{{\rm sin}(eFT)\over eFT}\biggr]
\nonumber\\
&&
\!\!\!\!\!\!\!\!\!\!\!\!\!\!\!\!\!\!\!\!\!\!\!\!\!\!\!\!\!
\!\!\!\!\!\!\!\!\!\!\!\!\!\!\!\!\!\!\!\!\!\!\!\!\!\!\!\!\!
\!\!\!\!\!
\times
\prod_{i=1}^N \int_0^T 
d\tau_i
\exp\biggl\lbrace\sum_{i,j=1}^N 
\bigl\lbrack \half k_i\cdot {\cal G}_{Bij}\cdot  k_j
-i\varepsilon_i\cdot\dot{\cal G}_{Bij}\cdot k_j
+\half
\varepsilon_i\cdot\ddot {\cal G}_{Bij}\cdot\varepsilon_j
\bigr\rbrack\biggr\rbrace
\mid_{\rm multi-linear}
\nonumber\\
\label{scalarqedmasterF}
\end{eqnarray}
\no
The determinant factor 
appearing here takes the change of the path integral
determinant due to the constant field
into account. Since by itself it just describes
the vacuum amplitude in a constant field one finds it
to be, of course, just the proper-time integrand of the
one-loop Euler-Heisenberg Lagrangian.

In \cite{adlsch}
this formalism has been applied
to a recalculation
of the QED photon splitting amplitude 
in a constant magnetic field $B$ \cite{adler}.
Using the modified master formula above with $N=3$
as a starting point
one finds, in the spinor QED case, the following
parameter integral representation for this
amplitude ($z=eBT$),

\begin{eqnarray}
&&C
[\omega,\omega_1,\omega_2,B] =
{m^8\over 4 \omega\omega_1\omega_2}
\int_0^{\infty}\!dT\,T{{\rm e}^{-m^2T}
\over {z}^2{\rm sinh}(z)}\nonumber\\
&&\times\int_0^T\!\!d\tau_1\,d\tau_2\,
\,{\rm exp}
\biggl\lbrace\!\!-{1\over 2}\!\sum_{i,j=0}^2\bar\omega_i
\bar\omega_j\Bigl[G_{Bij} + {T\over 2z}
{{\rm cosh}(z\dot G_{Bij})\over {\rm sinh}(z)}
\Bigr]\biggr\rbrace\nonumber\\
&&\times 
\Biggl\lbrace\biggl\lbrack
-\cosh (z)\ddot G_{B12}\!+\!\omega_1\omega_2
\Bigl( {\rm cosh}(z)-\cosh (z\dot G_{B12})\Bigr )\biggr\rbrack
\nonumber\\
&&\times\biggl\lbrack
{\omega\over\sinh(z)\!\cosh(z)}
-\omega_1
{{\rm cosh}(z\dot G_{B01})\over {\rm sinh}(z)}
-\omega_2
{{\rm cosh}(z\dot G_{B02})\over {\rm sinh}(z)}
\biggr\rbrack\nonumber\\
&&\!\! +{\omega\omega_1\omega_2
G_{F12}\over \cosh (z)}
\biggl\lbrack
\sinh (z\dot G_{B01})\Bigl (\cosh (z)\!-\!\cosh (z\dot G_{B02}) \Bigr )
- (1 \leftrightarrow 2) 
\biggr\rbrack \Biggl\rbrace
\non\\
\label{psspinresult}
\end{eqnarray}\no
Here 
$\omega$ ($\omega_{1,2}$)
denotes the energy of the incoming
(outgoing) photon(s),
and we have defined $\bar\omega_0=\omega,\bar\omega_{1,2}=-\omega_{1,2}$.
This integral formula is exact for arbitrary magnetic field
strengths and for photon energies up to the
pair creation threshold.
Of the various known integral representations for
this amplitude it is the most compact one.

\section[]{QED in a Constant Field (2 - Loop)}
\renewcommand{\theequation}{3.\arabic{equation}}
\setcounter{equation}{0}

Since the one - loop master formulas eqs.(\ref{scalarqedmaster}),
(\ref{scalarqedmasterF}) and their fermion loop generalizations
are valid off-shell, they can be used also for 
the construction of
higher - loop amplitudes \cite{ss3,rss}.
In \cite{rss} we obtained along these lines 
the following result for
the on-shell renormalized two - loop 
correction to the spinor QED
Euler-Heisenberg
Lagrangian in a constant magnetic field,

\bear
{\cal L}^{(2)}[B]
&=&
{\alpha\over
{(4\pi )}^{3}}
\int_{0}^{\infty}{dT\over T^3}e^{-m^2T}
\int_0^1 du_a
\,
\Bigl[
L(z,u_a,4)-L_{02}(z,u_a,4)-{g(z,4)\over G_{Bab}}
\Bigr]\nonumber\\
&&\!\!\!\!\!\!\!\!\!\!\!\!\!\!\!
\!\!\!\!\!\!\!\!\!\!\!\!\!\!\!
-
{\alpha\over {(4\pi)}^3}
m^2\int_0^{\infty}
{dT\over T^2}{\rm e}^{-m^2T}
\biggl[
{z\over\tanh(z)}
-{z^2\over 3}-1
\biggr]
\biggl[18-12\gamma-
12\ln (m^2T)+{12\over m^2T}
\biggr]
\nonumber\\
\label{Gammaspinrenorm}
\ear\no
where
$u_a \equiv {\mid\tau_a-\tau_b\mid\over T}$
and

\bear
G_{Bab}^z &\equiv& {T\over 2}
{\Bigl[\cosh (z)-\cosh(z\dot G_{Bab})\Bigr]\over
z\sinh (z)}\non\\
L(z,u_a,4)&=&{z\over\tanh(z)}
\Biggl\lbrace
B_1
{\ln ({G_{Bab}/G_{Bab}^z})
\over{(G_{Bab}-G_{Bab}^z)}^2}
+{B_2\over
G^z_{Bab}(G_{Bab}-G^z_{Bab})}
+{B_3\over
G_{Bab}(G_{Bab}-G^z_{Bab})}
\Biggr\rbrace\nonumber\\
B_1&=&4z
\Bigl(
\coth(z)-\tanh(z)
\Bigr)
G^z_{Bab}-4G_{Bab}
\quad  \nonumber\\
B_2&=&2\dot G_{Bab}\dot G^z_{Bab}+ 
z(8\tanh(z)-4\coth(z))
G^z_{Bab}-2
\quad \nonumber\\
B_3&=&4G_{Bab}
-2\dot G_{Bab}\dot G^z_{Bab}
-4z\tanh(z)G^z_{Bab}+2\nonumber\\
L_{02}(z,u_a,4)&=&-{12\over G_{Bab}}+2z^2,\quad
g(z,4)=-6\biggl[
{z^2\over{\sinh(z)}^2}+z\coth(z)-2
\biggr].
\label{defLLg}
\ear\no
Comparing with the two existing previous 
calculations 
of this amplitude we found agreement 
with \cite{ritusspin} 
to order of $O(B^{20})$ in the weak-field
expansion in $B$, and term by term agreement
with \cite{ditreu} after performance of an appropriate
finite electron mass renormalization.
This calculation was further generalized to the
case of a general constant field in \cite{ks}.

\section[]{Inclusion of Axial Vectors}
\renewcommand{\theequation}{4.\arabic{equation}}
\setcounter{equation}{0}

Finally, we mention the following recent
generalization of eq.(\ref{spinorpi})
to the case of a spinor loop coupled both to a vector
field $A$
and an axial vector field $A_5$
\cite{mcksch}:

\bear
\Gamma[A,A_5]
&=& -2\, \int_0^\infty \, 
\frac{dT}{T} 
\e^{-m^2T}
\Dx
\Dpsi
\, \e^
{
-\int_0^Td\tau\,
L(\tau)
}\\
L(\tau) &=&
\!\kinb
+\!\half\psi\cdot\dot\psi
+\!ie\dot x^{\mu}A_{\mu}
-\!ie\psi^{\mu}F_{\mu\nu}\psi^{\nu}
-\!2ie_5\hat\gamma_5\dot x^{\mu}\psi_{\mu}\psi_{\nu}A_5^{\nu}
+\!ie_5\hat\gamma_5\partial_{\mu}A^{\mu}_5
+\!2e_5^2A_5^2
\non\\
\label{vapi}\non
\ear\no
The periodicity properties of the Grassmann path integral 
$\int{\cal D}\psi$ are now
determined by the
operator $\hat\gamma_5$. After expansion of the interaction
exponential a given term in the integrand will have to be
evaluated using antiperiodic (periodic) boundary conditions
on $\psi$ 
if it contains $\hat\gamma_5$
at an even (odd) power. 
(After the boundary 
conditions are determined
$\hat\gamma_5$ can be replaced by unity.)
Thus for amplitudes involving an odd number of axial vectors
$\int{\cal D}\psi$ will have zero - modes, which
upon integration produce the expected $\varepsilon$ - tensor.
We have applied this path integral to a calculation
of the axial vacuum polarization tensor $<AA_5>$ in a general
constant field, with the result \cite{iorasc}

\be
\Pi_5^{\mu\nu}(k)
=
\half
\Tintm
\4piT4
\int_0^Td\tau_1\int_0^Td\tau_2
\,\,
J^{\mu\nu}
\e^{-Tk\cdot V \cdot k}
\label{axvpF}
\ee
\no
where
\bear
J^{\mu\nu} &=&
4iT\biggl\lbrace
\tilde F^{\mu\nu}kUk
+\Bigl[
(\tilde F k)^{\mu}(Uk)^{\nu}
+ (\mu\leftrightarrow\nu )
\Bigr]
\non\\
&&\quad
-(A_{12}\tilde F)^{\mu\nu}kA_{12}k
-\Bigl[
(\tilde F A_{12}k)^{\nu}(A_{12}k)^{\mu}
+ (\mu\leftrightarrow\nu )
\Bigr]
\biggr\rbrace
\non\\
&&
+T^2 F\cdot\tilde F
\biggl\lbrace
-S_{12}^{\mu\nu}kUk
-\Bigl[
(S_{12}k)^{\mu}(Uk)^{\nu}
+ (\mu\leftrightarrow\nu )
\Bigr]
\non\\
&&\quad
+(A_{12}S_{22})^{\mu\nu}kA_{12}k
+\Bigl[
(A_{12}k)^{\mu}(S_{22}A_{12}k)^{\nu}
+ (\mu\leftrightarrow\nu )
\Bigr]
\biggr\rbrace
\non\\
S_{12} &=&
i\Bigl(
{\cos(z\dot G_{B12})\over\sin(z)}
-{1\over z}
\Bigr)
\label{S12}, \,
S_{22} =
i\Bigl(
\cot(z)
-{1\over z}
\Bigr)
\label{S22},\,
A_{12} =
{\sin(z\dot G_{B12})
\over
\sin(z)}
 \non\\
U &=& {1-\cos(z\dot G_{B12})\cos(z)
   \over \sin^2(z)},\,
V = {\cos(z\dot G_{B12})-\cos(z)\over 2z\sin(z)}
\non\\
\label{S12etc}
\ear\no
($\tilde F$ denotes the dual field strength tensor).
For the magnetic special case this parameter integral
agrees with the known field theory result 
\cite{axvpft}. However in contrast to those field
theory calculations 
the path integral calculation is
manifestly gauge invariant.

\section[]{Summary}
\renewcommand{\theequation}{5.\arabic{equation}}
\setcounter{equation}{0}

From the examples given here it should be evident that
the string-inspired technique is an elegant and efficient
tool for the calculation of the QED photon S-matrix.
Some other quantities of interest which we hope to
obtain along these lines are the off-shell photon-photon
scattering amplitude, and gradient corrections to the
photon splitting amplitude. Also we are currently
working on a complete automatization of the calculation
of the QED effective action in the higher derivative
expansion \cite{flisch}.

\vfill\eject

\end{document}